\documentclass[pre]{revtex4}
\usepackage{graphics} 

\begin{document}  

\title{Coupled mode theory for photonic band-gap inhibition of spatial
instabilities}

\date{\today}

\author{Dami\`a Gomila}
\author{Gian-Luca Oppo}

\affiliation{Department of Physics, University of Strathclyde \\
107 Rottenrow, Glasgow, G4 0NG, United Kingdom}

\begin{abstract}  

We study the inhibition of pattern formation in nonlinear optical systems 
using intracavity photonic crystals. We consider mean field models for single
and doubly degenerate optical parametric oscillators. Analytical expressions
for the new (higher) modulational thresholds and the size of the ''band-gap``
as function of the system and photonic crystal parameters are obtained via a
coupled-mode theory. Then, by means of a nonlinear analysis, we derive
amplitude equations for the unstable modes and find the stationary solutions
above threshold. The form of the unstable mode is different in the lower and
upper part of the bandgap. In each part there is bistability between two 
spatially shifted patterns. In large systems stable wall defects between the
two solutions are formed and we provide analytical expressions for their shape.
The analytical results are favorably compared with results obtained from the
full system equations. Inhibition of pattern formation can be used to spatially
control signal generation in the transverse plane.

\end{abstract}


\maketitle

\section{Introduction}

Photonic crystals (PC) have shown to be able to control light in ways that were
not possible with conventional optics \cite{Joannopoulos,Knight}. Their
remarkable properties stem from the unusual dispersion relation as a result of 
the periodic modulation of their dielectric properties. Most of the interest on
PC is related to propagation problems. Here the existence of photonic
band-gaps, i.e. a range of frequencies for which  light can not propagate in
the medium, allows for a new way of guiding and localize light once defects are
introduced in the PC \cite{Joannopoulos,Knight}. Photonic crystals in
combination with  nonlinear effects have been also considered for all-optical 
switching devices \cite{Slusher}. 

Transverse effects in periodic media have been mainly studied in propagation in
planar waveguides with periodic modulation of the refractive index in the
transverse direction and arrays of couple waveguides. In the first case the
attention has been focused on the so called spatial gap (or Bragg) solitons
\cite{Kivshar,Nabiev,Yulin}, which are intense peaks with frequencies inside
the (linear) band-gap. This is possible thanks to a shift of the photonic
band-gap  boundaries due to nonlinear effects: the intense core of a gap
soliton can propagate freely in the periodic media, while in its less intense
tails, where non-linearity can be neglected, light is reflected back to the
center due to Bragg reflection, sustaining the localized structure. Bragg
solitons are usually studied within the coupled-mode theory where only the
slowly varying envelope of two counter-propagating beams are considered. Arrays
of waveguides are usually studied within the tight-binding approximation, where
the system is described by a set of coupled ordinary  differential equations
describing the dynamics of the guided-mode amplitude in each site. The
evanescent coupling between adjacent waveguide give rise to the so called
discrete diffraction. If nonlinearity is also present discrete solitons can be
formed \cite{Slusher,Kivshar}.

In this paper we consider a different case, a photonic crystal inside a
nonlinear optical cavity, i.e. a system with driving and dissipation. Nonlinear
optical cavities typically undergo spatial instabilities leading to the
formation of spatial structures similar to what observed in many different 
fields across science \cite{Cross-Hohenberg,Walgraef}. In particular, spatial 
structures in nonlinear optical cavities have important potential applications 
in photonics such as memories, multiplexing, optical processing and imaging
\cite{Firth}. Control of spatial structures is then an important issue for the
implementation of such devices. In this context, different mechanisms of
control have been proposed \cite{Martin,Harkness,Neubecker}. In particular,
Neubecker and Zimmerman reported on experimental pattern formation in presence
of an external modulated forcing and observed lockings between the forcing and
natural wavelengths \cite{Neubecker}. More recently the use of the properties
of photonic crystals for the control of optical spatial structures  has been
considered  \cite{PRL,Dabbicco,Choquette}. In \cite{PRL} we showed how the
photonic band-gap of a photonic crystal could be used to inhibit a pattern
forming instability in a self-focusing Kerr cavity. Pattern formation in an
optically-injected, photonic-crystal vertical-cavity surface-emitting laser,
electrically biased below threshold has been observed experimentally in
\cite{Dabbicco}. Finally, in \cite{Choquette}, a defect in a photonic crystal 
has been used to achieve stable emission in broad area lasers.

Here we study the inhibition of pattern formation in a mean field model for a
Degenerate Optical Parametric Oscillator with a photonic crystal. We first show
that the inhibition mechanism introduced in \cite{PRL} occurs independently of
the type of nonlinearity here being quadratic. Then we introduce an  analytical
treatment by means of a couple-mode theory that fully explains the phenomenon.
A linear stability analysis allows us to determine analytically the new
thresholds shifted in parameter space by pattern inhibition and the form of the
unstable modes. We show  that the band-gap region is divided into two different
regions: one in which pattern formation takes place due to an energy
concentration at the maxima of the photonic crystal modulation, and a different
one where the energy concentrates at the minima. We determine amplitude
equations for the unstable modes and find the stationary solutions above
threshold analytically. We show  that this scenario is organized by a
co-dimension two bifurcation point where both modes become simultaneously
unstable. We also show that in the band-gap, where the pattern arises with a
wavenumber half the one of the PC, there is bistability between two different 
"phase locked" solutions differing by a shift of a PC wavelength in the near
field. This bistability stems from the breaking of the translational symmetry
by the PC. In large systems this leads to the formation of domains of different
"phases" connected by topological walls (or antiphase  boundaries)
\cite{Walgraef}. This is a different kind of (spatial) bistability where the
two different "phases" do not differ on the phase of the electric field but on
the location of the spatial oscillation in the near field.

\section{Models}

\begin{figure} 
\includegraphics{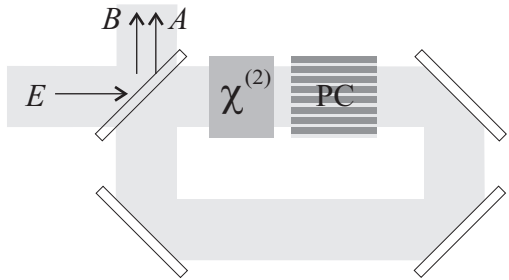}
\caption{\label{cavity} Scheme of a doubly resonant degenerate optical
parametric oscillator with a photonic crystal. It consist of a ring cavity
filled with a  quadratic ($\chi^{(2)}$) medium and a photonic crystal slab
(PC). $E$ is the plane wave input field at frequency $\omega$ partially
transmitted into the cavity. The other mirrors are assumed to be perfectly
reflecting. $B$ and $A$ are the pump and signal fields at frequencies $\omega$
and $\omega/2$ respectively. In the single resonant case only the signal field
$A$ resonates with the cavity.} 
\end{figure}

A phase-matched Double Resonant Degenerate Optical Parametric Oscillator 
(DRDOPO), where both pump and signal fields resonate in the cavity, can be
described in the mean field approximation by \cite{refsDRDOPO}:
\begin{eqnarray}
\label{DRDOPO} 
\partial_t B &=& \Gamma [ -B + E - A^2 ] + 
i {1 \over 2} \partial_{xx} B \nonumber \\ 
\partial_t A &=& -A - i\Delta_s A + BA^* + i \partial_{xx} A, 
\end{eqnarray} 
where $B$ and $A$ are the pump and signal slowly varying amplitudes,  $\Gamma$
is the ratio  between the pump and signal cavity decay rates and $E$ is the
amplitude of  the external pump field (our control parameter). $\Delta_s=\Delta
+ \alpha f(x)$ is the total signal detuning, where $\Delta$ is the average
detuning and $f(x)=e^{ik_{pc}x}+e^{-ik_{pc}x}$ describes the spatially
dependent contribution of the photonic crystal (Fig.~\ref{cavity}).

For the singly resonant DOPO (SRDOPO), where there is no cavity for the pump
field, the mean field equation for the resonating signal is 
\cite{refsSRDOPO,G-L}
\begin{equation}
\label{SRDOPO} 
\partial_t A = -A - i\Delta_s A + E A^* -|A|^2A+ 
i \partial_{xx} A.
\end{equation} 

\section{Linear stability analysis with periodic media: couple-mode
theory}

The linearization around the steady state homogeneous solution $B^0=E$ and 
$A^0=0$ of Eqs.~(\ref{DRDOPO}) and (\ref{SRDOPO}) leads, in both cases, to the
same equation for the perturbations of the signal $A$:
\begin{equation}
\label{linear}
\partial_t A = -(1 + i\Delta) A -i \alpha f(x) A + E A^* + \partial_{xx} A. 
\end{equation} 
Without the photonic crystal ($\alpha =0$), the homogeneous solution is stable 
for $E<1$. Above threshold ($E>1$) a stripe pattern arises with a wavenumber
$k_c=\sqrt{-\Delta}$. Here we restrict ourselves to negative values of the
detuning. For positive detuning the system display bistability between
homogeneous solutions \cite{G-L}. For $\alpha \ne 0$, and assuming that  the
amplitude of the modulation is weak enough, we can write the signal
perturbations as a superposition of two waves with opposite transverse
wavenumbers \cite{Yulin}
\begin{equation}
\label{ansatz}
A=A_+(x,t)e^{ik_{pc}x/2}+A_-(x,t)e^{-ik_{pc}x/2},
\end{equation}
where $A_\pm(x,y)$ are slow functions of $x$. This is basically equivalent to
the so called {\it coupled-mode theory} for propagation of pulses in periodic
media \cite{Slusher}.

By writing $A_\pm=a_\pm^+(t)e^{ikx}+a_\pm^-(t)e^{-ikx}$, one obtains  a set of
coupled linear ordinary differential equations for the amplitudes  $a_\pm^+$
($a_\pm^-$)  of the  Fourier components $\pm k_{pc}/2 + k$ ($\pm k_{pc}/2 - k$)
of the  perturbations. By defining ${\bf F}=(a_+^+,a_-^{-*},a_-^+,a_+^{-*})^T$
this set of equations can be written as 
\begin{equation}
\label{linear set}
\dot{\bf F}={\mathcal L}{\bf F},
\end{equation}
where ${\mathcal L}$ is given by
\begin{equation}
\label{J}
{\mathcal L}= \left( \begin{array}{cccc} 
-1-i\Delta-i(k+{k_{pc} \over 2})^2 & E & -i\alpha & 0 \\ 
E & -1+i\Delta+i(k+{k_{pc} \over 2})^2 & 0 & i\alpha \\ 
-i\alpha & 0 & -1-i\Delta-i(k-{k_{pc} \over 2})^2 & E \\ 
0  & i\alpha & E & -1+i\Delta + i(k-{k_{pc} \over 2})^2 
\end{array} \right).
\end{equation} 
In this way one reduces the stability analysis of the steady state of the
initial partial differential equations with periodic coefficients
(\ref{DRDOPO}) and (\ref{SRDOPO}) to diagonalize the $4\times 4$ complex 
matrix ${\mathcal L}$. In the case $k=0$ one has to consider that 
$a_\pm^+=a_\pm^-$.

The eigenvalue of ${\mathcal L}$ with largest real part is 
\begin{eqnarray}
\label{eigenvalues}
\lambda = -1+\sqrt{E^2-\left[k^2+\left({k_{pc} \over 2}\right)^2 + \Delta 
- \sqrt{\alpha^2 + 4k^2\left({k_{pc} \over 2}\right)^2}\right]^2},
\end{eqnarray}
and by setting $\lambda=0$ we obtain four marginal stability curves
\begin{eqnarray}
\label{marginal}
\Delta_1(k,E)&=&-d_1(k)+ d_2(k)+ d_3(E) \nonumber \\ 
\Delta_2(k,E)&=&-d_1(k)+ d_2(k)- d_3(E) \nonumber \\
\Delta_3(k,E)&=&-d_1(k)- d_2(k)+ d_3(E) \nonumber \\
\Delta_4(k,E)&=&-d_1(k)- d_2(k) - d_3(E).
\end{eqnarray}
where $d_1(k)= \left(k_{pc} / 2\right)^2 + k^2 $, $d_2(k)= \sqrt{4 k^2
\left({k_{pc}/ 2}\right)^2 + \alpha^2} $ and $d_3(E) = \sqrt{E^2-1}$.

\begin{figure} 
\includegraphics{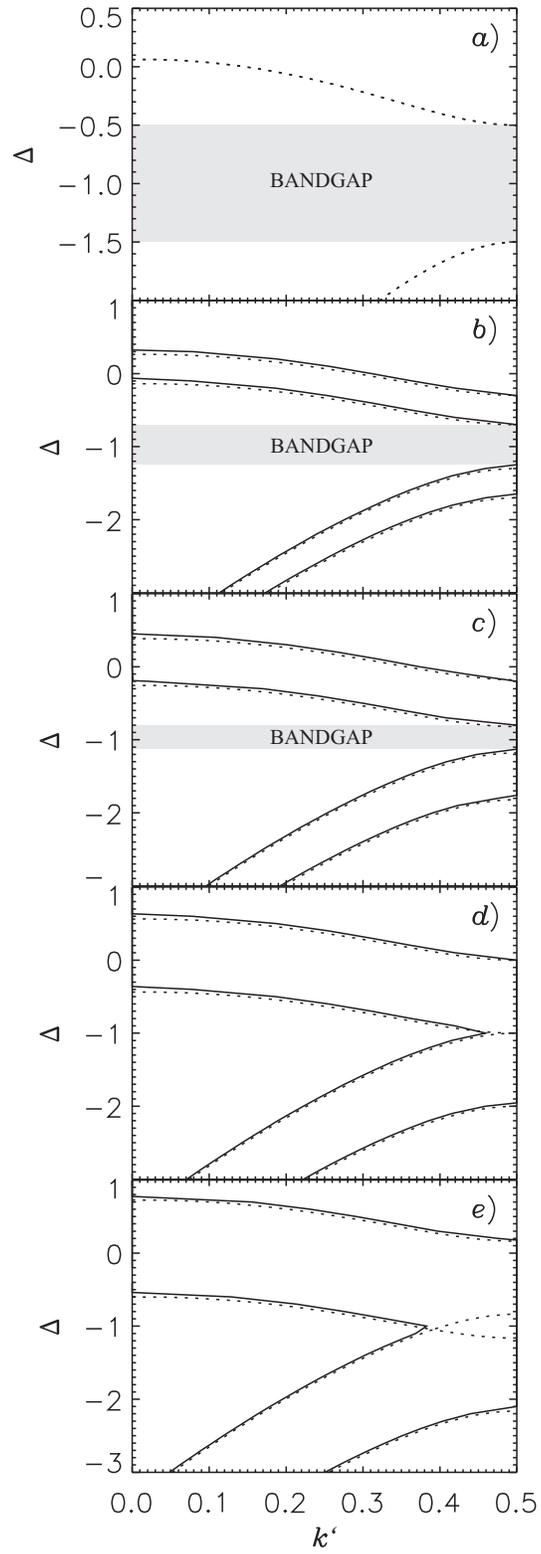}
\caption{\label{marginal1} Unstable wavenumbers of a DOPO for $a) E=1.0$, $b)
E=1.02$, $c) E=1.05$, $d) E=\sqrt{1+\alpha^2}=1.118034$, and $e) E=1.2$ in
presence of a periodically modulated media ($k_{pc}=2, \alpha=0.5$). For each 
value of $\Delta$ wavenumbers between the two lines are  unstable.} 
\end{figure}

Fig.~\ref{marginal1} shows the marginal stability curves for different values
of the pump $E$. Note that, due to definition (\ref{ansatz}), $k=0$ indicates a
perturbation with a wavenumber at the limit of the first Brillouin zone.
Therefore it is convenient to plot $\Delta$ as a function of
$k'=(k_{pc}/2-k)/k_{pc}$. Dotted lines are the results from the coupled-mode
theory (\ref{marginal}), while solid lines have been obtained from a numerical
stability analysis of the full model, i.e. solving the eigenvalue problem
associated to the linear differential operator with periodic coefficients in
the rhs of (\ref{linear}) \cite{PRL}. The coupled-mode theory provides a very
good analytical approximation for thresholds and unstable wavenumbers allowing
us to predict the existence and size of a band-gap in the modulation
instability analytically.  In the following we analyze the results of the
coupled-mode theory in more  detail.

From (\ref{marginal}) one can evaluate the instability threshold of the
fundamental solution as a function of the detuning $\Delta$. For $E<1.0$ the
steady state homogeneous solution is stable for all values of the detuning. At
$E=1$ (Fig.~\ref{marginal1}a) $d_3(E)=0$ and the four marginal stability lines
given by Eqs.~(\ref{marginal}) become only two,  signaling the instability
threshold for all values of the detuning outside the band-gap ($-d_1(k_{pc}/2)
- d_2(k_{pc}/2) + \alpha^2 < \Delta <  - (k_{pc}/2)^2 - \alpha < -(k_{pc}/2)^2
+ \alpha  < \Delta <  - d_1(k_{pc}/2) + d_2(k_{pc}/2)+\alpha^2$). In the
bandgap,  $-(k_{pc}/2)^2-\alpha < \Delta < -(k_{pc}/2)^2+\alpha$, the 
instability is, however, inhibited. The threshold $E=1$ for values of $\Delta$
outside the bandgap is slightly underestimated. In the full model,  for this
value of the pump, the system is still stable (no solid line in
Fig.\ref{marginal1}a). This is due to the fact that the spatial modulation
couples the fundamental wavenumbers with their harmonics, which are dumped,
introducing an additional source of stability. Harmonics are not taken into
account in the couple-mode theory and then the threshold is slightly lower than
for the full models (\ref{DRDOPO}) and (\ref{SRDOPO}). 

Increasing further the value of the pump the band-gap narrows
(Figs~\ref{marginal1}b,c). At $E=\sqrt{1+\alpha^2}$ the homogeneous solution
become eventually unstable for  any value of the detuning
(Fig.~\ref{marginal1}d). The threshold value of $E$ as a function of the
photonic crystal parameters for any value of $\Delta$ is given by
\begin{equation}
\label{thresholdtheo}
E^2_{th}(\Delta)= \left\{\begin{array}{ccccc} 
1 & & -({k_{pc}\over 2})^2+\alpha &< \Delta \le&
-d_1({k_{pc}\over 2}) + d_2({k_{pc}\over 2})+\alpha^2 \\
1+\left[-({k_{pc}\over 2})^2+\alpha-\Delta\right]^2 & &
  -({k_{pc}\over 2})^2 &< \Delta \le& -({k_{pc}\over 2})^2+\alpha \\
1+\left[({k_{pc}\over 2})^2+\alpha+\Delta\right]^2 & &
  -({k_{pc}\over 2})^2-\alpha &< \Delta \le& -({k_{pc}\over 2})^2 \\
1 & & -d_1({k_{pc}\over 2}) - d_2({k_{pc}\over 2}) + \alpha^2 & < 
\Delta \le& -({k_{pc}\over 2})^2-\alpha  
\end{array} \right.,
\end{equation}
and the critical wavenumber by:
\begin{equation}
\label{kcoutside}
k^2_c(\Delta)= \left\{ \begin{array}{ccccc} 
(k_{pc}/ 2)^2-\Delta - \sqrt{-4({k_{pc}\over 2})^2 \Delta + \alpha^2} & &
-({k_{pc}\over 2})^2 + \alpha &< \Delta \le&
-d_1({k_{pc}\over 2}) + d_2({k_{pc}\over 2})+\alpha^2  \\
0 & & -({k_{pc}\over 2})^2 &< \Delta \le& -({k_{pc}\over 2})^2+\alpha \\
0 & & -({k_{pc}\over 2})^2-\alpha &< \Delta \le&  -({k_{pc}\over 2})^2 \\
({k_{pc}\over 2})^2-\Delta - \sqrt{-4({k_{pc}\over 2})^2 \Delta + \alpha^2} & &
-d_1({k_{pc}\over 2})^2 - d_2({k_{pc}\over 2}) + \alpha^2 &< 
\Delta \le& -({k_{pc}\over 2})^2-\alpha  
\end{array} \right..
\end{equation}

\begin{figure} 
\includegraphics{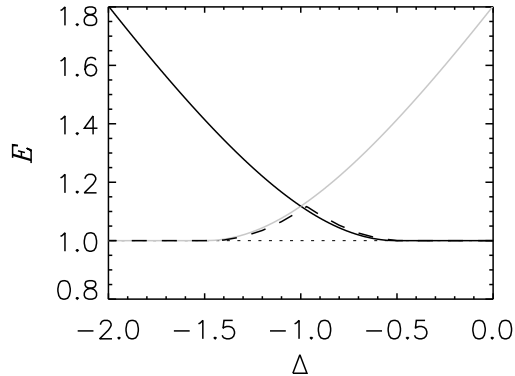}
\caption{\label{threshold} Pump threshold as function of the detuning from the
theoretical result (\ref{thresholdtheo}) obtained by means of the coupled-mode
theory (solid black and grey lines). The dashed line is the result obtained
from a numerical stability analysis of the full model. The dotted line is the
threshold for a DOPO without photonic crystal.} 
\end{figure}

Fig.~\ref{threshold} shows the theoretical prediction (\ref{thresholdtheo}) for
the instability threshold of the homogeneous solution (solid lines). The black
and grey lines correspond to two different eigenmodes. The first is a mode with
energy concentrated in the minima of the photonic crystal, while the second in
the maxima (see next section for more details). The  dashed line is the
threshold for the full system computed as explained above. 

The coupled-mode theory accurately capture the features of the inhibition of
the modulation instability by the photonic crystal. The fact that the gap of
values of the detuning for which pattern formation is inhibited decreases with
the intensity of the pump is because nonlinearity always overcomes the linear
inhibition by the photonic crystal for suitable high input intensities. This is
similar to the phenomenon behind the formation of gap solitons. But in that
case the nonlinearity only shifts the position of the band-gap, while  in our
case it narrows it and makes it eventually to disappear.

\begin{figure} 
\includegraphics{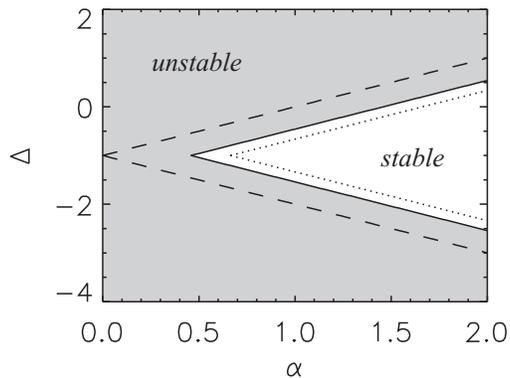}
\caption{\label{stabdiag} Stability diagram of the steady state homogeneous
solution for $E=1.1$ (solid line). The system is unstable in the shadowed 
region. The dashed (dotted) line shows the boundary of the stable region for 
$E=1.0$ ($E=1.2$).} 
\end{figure}

From Eq.~(\ref{thresholdtheo}) we can calculate the size $g(E,\alpha)$ of the 
band-gap as function of the system and photonic crystal parameters:
\begin{equation}
\label{band-gap}
g(E,\alpha)=2(\alpha-\sqrt{E^2-1}).
\end{equation}
The size of the band-gap is proportional to the amplitude of the modulation
$\alpha$ and becomes smaller by increasing the pump $E$ (Fig.~\ref{marginal1}),
eventually disappearing for $E=\sqrt{1+\alpha^2}$  (Fig.~\ref{marginal1}d).
For  $E>1+\alpha^2$ the homogeneous solution is unstable for any value of
$\Delta$  (Fig.~\ref{marginal1}e). Fig.~\ref{stabdiag} shows how the width of
the band-gap (white region) changes as function of the amplitude of the
modulation $\alpha$ for a fixed value of the pump $E$. The presence of the
modulated medium opens  an entirely new stable region. Fig.~\ref{stabdiag}
should be compared with Fig.~6 of \cite{PRL} showing that pattern inhibition is
independent of the form of the nonlinearity.

\begin{figure} 
\includegraphics{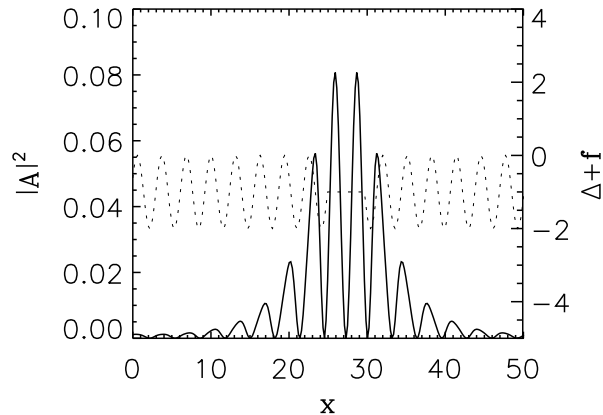}
\caption{\label{defectls} Localized spot of signal generation due to a defect
in the photonic crystal. Here $E=1.08, \Delta=-1, k_{pc}=2$ and $\alpha=0.5$.} 
\end{figure}

Since for negative signal detunings, in the absence of PC, down conversion
takes place at a finite wavenumber, the inhibition mechanism can be used to
spatially control the generation of signal in a way analogous to that explained
in \cite{PRL}. If we set our system in a parameter region where pattern
formation is inhibited, the inclusion of a defect in the PC will lead to a spot
of signal generation as shown in Fig.~\ref{defectls}. 

\section{Weakly nonlinear analysis: amplitude equations for pattern formation in
periodic media}

In this section we study by means of a multiple scales analysis the solutions
that appear above threshold for values of the detuning inside the band-gap. 
Above threshold nonlinear terms have to be considered since they  saturate the
linear growth induced by the instability. In the following we consider the
SRDOPO only. By including the  nonlinear terms from (\ref{SRDOPO}) in
(\ref{linear set}) we obtain: 
\begin{equation}
\label{NL coupled mode}
\dot{\bf F}={\mathcal L}{\bf F} - {\mathcal W}({\bf F}),
\end{equation}
where ${\mathcal W}=[(a_+^+ a_+^{-*} + 2 a_-^{-*}a_-^+) a_+^+ , (2 a_-^+
a_-^{-*} +  a_+^{-*}a_+^+) a_-^{-*} (a_-^+ a_-^{-*} + 2 a_+^{-*}a_+^+) a_-^+,
(2 a_+^+ a_+^{-*} +  a_-^{-*}a_-^+) a_+^{-*}]^T$ is a nonlinear function of 
${\bf F}$.

In the bandgap, the critical wave number is always $k_c=0$ independently of the
detuning. We recall that in this case one has to consider $a_\mp^+ = a_\mp^-$. 
In the following we will discuss the lower ($-(k_{pc}/2)^2-\alpha \leq \Delta
<  -(k_{pc}/2)^2$) and upper ($-(k_{pc}/2)^2 \leq \Delta < -(k_{pc}/2)^2+ 
\alpha$) halves of the band-gap separately.

\subsection{Lower-half part}
Assuming the following scaling \cite{refsDRDOPO}:
\begin{eqnarray}
\label{scaling}
{\bf F} &=& \epsilon {\bf F_1}+\epsilon^3 {\bf F_3} \\ \nonumber
E &=& E_{th}+\epsilon^2 E_2 \\ \nonumber
T &=& \epsilon^2 t \\ \nonumber
X &=& \epsilon^2 x.
\end{eqnarray}
Substituting  (\ref{scaling}) in (\ref{NL coupled mode}), at order $\epsilon$
one obtains ${\bf F_1}=F_1 {\bf v_1}$, where ${\bf v_1}$ is the critical
eigenmode of ${\mathcal L}$   (${\mathcal L} {\bf v_1}=0$ and $|{\bf v_1}|=1$)
and $F_1$ is its real amplitude.  In the near field, ${\bf v_1}$ has the form: 
$[(1+E_{th}-i \sqrt{E_{th}^2-1})/\sqrt{2E_{th}(1+E_{th})}]\cos(k_{pc}x/2)$. 

The solvability condition at order $\epsilon^3$ yields to the following equation
for the real amplitude $\tilde{A_1}=\epsilon F_1$ of the unstable mode
\begin{equation}
\label{amplitudeq1}
\partial_t \tilde{A_1}=\sqrt{E_{th}^2-1} ~ \partial_x^2 \tilde{A_1}
+ E_{th}(E-E_{th})\tilde{A_1} - {3 \over 4} \tilde{A_1}^3.
\end{equation}
The amplitude equation (\ref{amplitudeq1}) is equivalent to the one obtained 
from a secondary instability at twice the spatial period of a cellular pattern 
\cite{Coullet}, except for the fact that in our case the translational 
invariance has been broken by the presence of the photonic crystal. The
homogeneous steady state solution of (\ref{amplitudeq1}) is $\tilde{A_1}=\pm
2\sqrt{E_{th}(E-E_{th})/3}$. The solution of (\ref{SRDOPO}) is then:
\begin{equation}
\label{patternsol1}
A(x)=\pm \sqrt{2 \over 3} \sqrt{E-E_{th} \over 1+E_{th}} 
\left( 1+E_{th}-i\sqrt{E_{th}^2-1} \right) \cos({k_{pc} \over 2}x).
\end{equation}

\begin{figure} 
\includegraphics{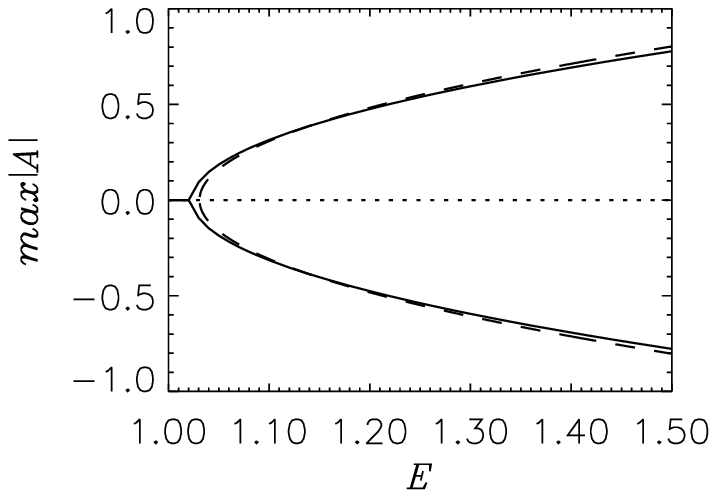}
\includegraphics{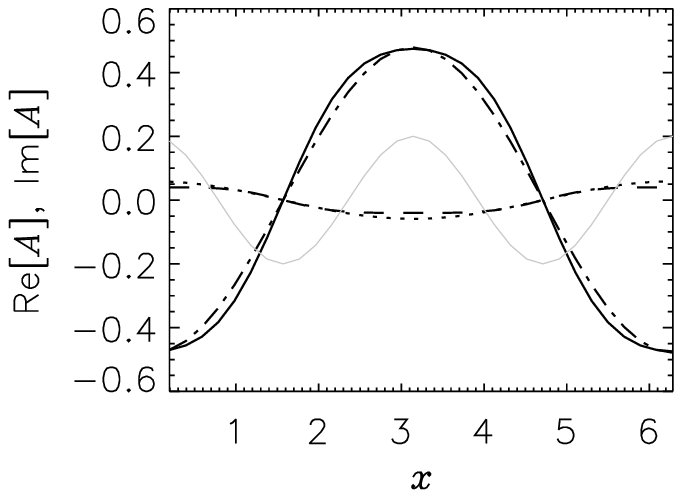}
\caption{\label{bifurcationcos} Left: Bifurcation diagram for the pattern
arising above threshold in the lower-half of the band-gap. Here $\Delta=-1.2$.
For this value of the detuning $E_{th}=1.04403$. Right: Real (solid line) and
imaginary (dashed lined) of the pattern solution of the full model for $E=1.2$.
The dot-dashed and dotted lines correspond to the real and imaginary parts of
the pattern from the coupled-mode theory. The grey solid line illustrates the
modulation of the photonic-crystal for comparison.}
\end{figure}

\begin{figure} 
\includegraphics{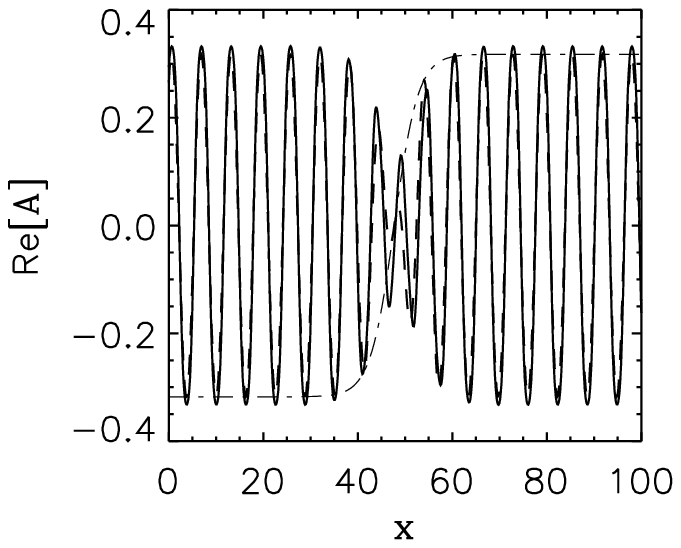}
\includegraphics{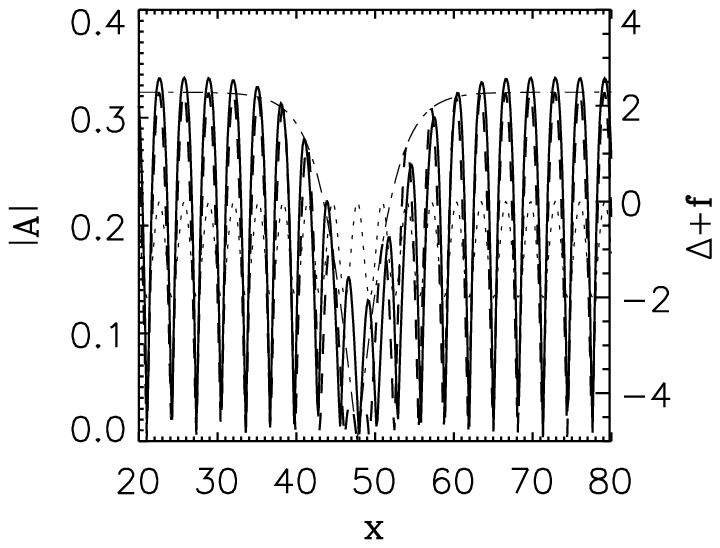}
\caption{\label{domainwall1} Two domains corresponding to the plus and minus
sign solutions (\ref{patternsol1}) separated by domain walls. The final state
is the result of a numerical simulation starting from an arbitrary initial 
condition. Left: Real part of the field. Right: Close up to the intensity
around a domain wall. The dotted line shows the modulation of the photonic
crystal.  Here $\Delta=-1.1$, $E=1.15$, $\alpha=0.5$ and $k_{pc}=2.0$.} 
\end{figure}

The bifurcation diagram and spatial form of this solution is shown in
Fig~\ref{bifurcationcos}. The plus and minus solutions (\ref{patternsol1}) are
created in a pitchfork bifurcation. The analytical solution (\ref{patternsol1})
is in very good agreement with the stationary solution of the full model
computed numerically. Note that, despite being completely equivalent, the plus
and minus solutions of (\ref{patternsol1}) are not the same. One corresponds to
the other shifted by a wavelength of the photonic crystal modulation. In a
system with translational invariance the position of a solution that brakes
such symmetry is undetermined. The photonic crystal periodicity selects just
two of a continuum of possible solutions. This two "frequency locked" solutions
differ now by a shift of a photonic crystal wavelength in the transverse
position. This situation is the spatial analogue of an oscillatory system
forced at twice its natural frequency \cite{Coullet2}. In a large system and
starting from arbitrary initial conditions, some sections of the system will
attain the first solution while others will move to the second one, leading to
the formation of domain walls.  This is illustrated in Fig.~\ref{domainwall1}.
The solid line is the result of a simulation of the full model (\ref{SRDOPO}) 
starting from a random initial condition. The dashed line is the steady state 
solution of Eq.~(\ref{amplitudeq1}) connecting the two (plus and minus) 
homogeneous solutions \cite{Walgraef} times the critical eigenmode  ${\bf v_1}$
in the near field:
\begin{equation}
\label{tanh1}
A_1(x)=\sqrt{2 \over 3}\sqrt{E-E_{th}\over 1+E_{th}} 
\left( 1+E_{th}-i\sqrt{E_{th}^2-1} \right)
\tanh \left[ \sqrt{E_{th}(E-E_{th}) \over 2} {x\over (E_{th}^2-1)^{1\over 4}}
\right]\cos({k_{pc} \over 2}x).
\end{equation}
The dotted-dashed line in Fig.~\ref{domainwall1} is the envelope of 
(\ref{tanh1}). The analytical result (\ref{tanh1}) is in very good agreement
with the domain wall obtained from the numerical simulations of the full model
(\ref{SRDOPO}).

\subsection{Upper-half part} 

In the upper-half part of the photonic band-gap the critical mode ${\bf v_1}$ 
has the form: $\{[\sqrt{E_{th}^2-1}+i(E_{th}-1)]/\sqrt{2E_{th}(E_{th}-1)}\} 
\sin(k_{pc}x/2)$. As in the previous case, we obtain a similar amplitude 
equation for the real amplitude $\tilde{A_1}$ of the unstable mode
\begin{equation}
\label{amplitudeq2}
\partial_t \tilde{A_1}=\sqrt{E_{th}^2-1} ~ \partial_x^2 \tilde{A_1}
+ \sqrt{E-E_{th}}\tilde{A_1} - {3 \over 4} \tilde{A_1}^3.
\end{equation}
In this case the pattern solutions is
\begin{equation}
\label{patternsol2}
A_1(x)=\pm  \sqrt{2 \over 3} 
\sqrt{E-E_{th} \over E_{th}(E_{th}-1)} 
\left[ \sqrt{E_{th}^2-1}+i(E_{th}-1) \right] \sin({k_{pc} \over 2}x).
\end{equation}

\begin{figure} 
\includegraphics{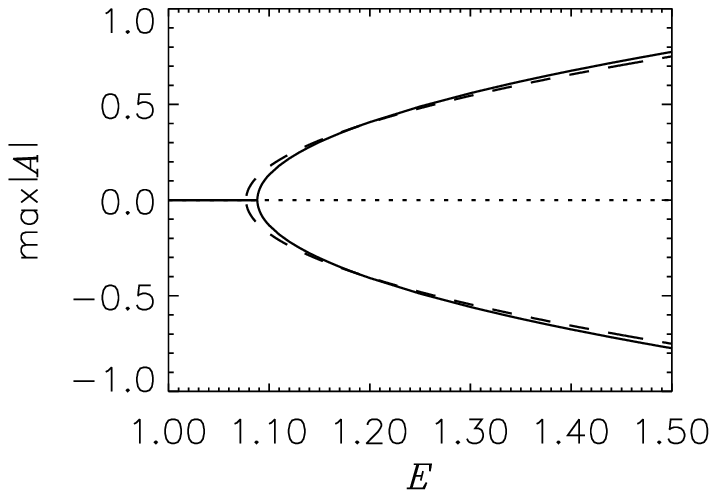}
\includegraphics{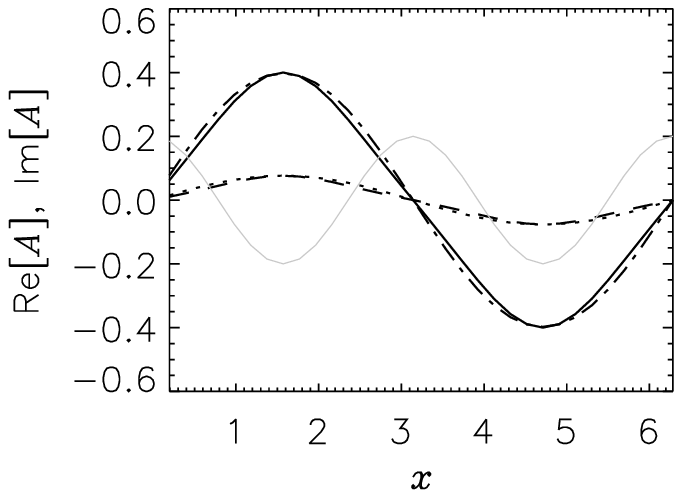}
\caption{\label{bifurcationsin} Left: Bifurcation diagram for the pattern
arising above threshold in the lower-half of the band-gap. Here $\Delta=-0.9$.
For this value of the detuning $E_{th}=1.07703$. Right: Real (solid line) and 
imaginary (dashed lined) of the pattern solution of the full model for $E=1.2$.
The dot-dashed and dotted lines correspond to the real and imaginary parts of
the pattern from the coupled-mode theory. The grey solid line illustrates the
modulation of the photonic-crystal for comparison.}
\end{figure}
The bifurcation diagram and spatial form of this solution is shown in
Fig~\ref{bifurcationsin}. The analytical solution (\ref{patternsol2}) is in
very good agreement with the stationary solution of the full model computed
numerically. As in the previous case, in large systems, fronts between the 
plus and minus solutions are formed. The shape of the front is given by:
\begin{equation}
\label{tanh2}
A_1(x)=2\sqrt{2 \over 3}\sqrt{E-E_{th}\over E_{th}(E_{th}-1)} 
\left[ \sqrt{E_{th}^2-1}+i(E_{th}-1) \right]
\tanh \left[ \sqrt{E-E_{th} \over 2} {x\over (E_{th}^2-1)^{1\over 4}}
\right]\sin({k_{pc} \over 2}x).
\end{equation}

Note that while the cosine solution in the lower-half part of the band-gap has
the maxima of the intensity at the maxima of the photonic crystal modulation,
the sine solution in the upper-part  has them at the minima. This different
distribution of energy in the photonic crystal is at the basis of the creation
of the band-gap. An interesting point is the middle of the band-gap
($\Delta=-(k_{pc}/2)^2$). At this particular value of the detuning, for
$E=E_{th}=\sqrt{1+\alpha^2}$ both the sine and cosine modes become
simultaneously unstable. This is a co-dimension two point where Eqs.
(\ref{amplitudeq1}) and (\ref{amplitudeq2}) become coupled. The unfolding of
such a critical point is however beyond the scope of this paper and it is left 
for future investigation. 

\section{Conclusions} 

In this paper we have studied pattern formation in nonlinear optical cavities
in presence of a photonic crystal, i.e. a spatial modulation of the refractive 
index. The linear phenomenon of the band-gap inhibits pattern formation for a
certain range of cavity detunings (bandgap). For high enough intensities
nonlinearity finally overcomes the inhibition by the photonic crystal and a
pattern arises. By means of a couple-mode theory approach we have obtained
analytical expressions for the new (shifted) threshold and the form of the
unstable modes. The bandgap is naturally divided in two halves, the lower-half
one in which the unstable mode has a cosine shape, i.e. its intensity maximums
are in correspondence with the maximums of the photonic crystal modulation, and
the upper-half where the unstable mode is a sine; the intensity maximums are in
correspondence with the minima of the photonic crystal modulation. By means of
a multiple scale analysis we also found the pattern solution above threshold.
In each part of the bandgap there is bistability between the plus and minus
cosine or sine solutions. This bistability stems from the breaking of the
translational symmetry of the photonic crystal. In large systems domain wall
between this two solutions are typically form. The shape of the defect wall is
given by an hyperbolic tangent. While the particular form of the coefficients
are model dependent, the shape of the unstable modes and the splitting of the
bandgap in two different regions are generic. We have checked that the same
phenomenon is present in a completely different model, namely the Kerr cavity
model studied in \cite{PRL}. Finally we also have shown that photonic crystal
can be useful to engineer particular spatial signal outputs in frequency down
conversion. An interesting extension of this work is to consider the case with
two transverse dimension. In this case, some of the features studied here, such
us the two different modes in each part of the bandgap, will remain the same,
while new features like the coupling of the geometry of the photonic crystal 
and that of the spontaneous pattern will come into play. 

\begin{acknowledgments} 
We thank A.J. Scroggie for useful discussions. We acknowledge  financial
support from EPSRC (GR S28600/1 and GR R04096/01), SGI  the Royal Society -
Leverhulme Trust, and the European Commission  (FunFACS).
\end{acknowledgments}

\end{document}